\documentstyle[aps,twocolumn,epsfig,floats]{revtex}

\begin{document}  
\draft

\wideabs{  
\author{Christof Gattringer, P.E.L.~Rakow, 
Andreas Sch\"afer and Wolfgang S\"oldner}
\address{\medskip Institut f\"ur Theoretische Physik, Universit\"at
Regensburg, D-93040 Regensburg, Germany} 
\date{February 7, 2002}
\title{Chiral symmetry restoration and the Z$_3$ sectors of QCD} 
\maketitle
   
\begin{abstract}  
Quenched SU(3) lattice gauge theory shows three phase transitions,
namely the chiral, the deconfinement and the Z$_3$ phase transition.
Knowing whether or not the chiral and the deconfinement phase transition 
occur at the same temperature for all Z$_3$ sectors could
be crucial to understand the underlying microscopic dynamics.
We use the existence of a gap in the Dirac spectrum as an order
parameter for the restoration of chiral symmetry.
We find that the spectral gap opens up 
at the same critical temperature in all Z$_3$ sectors in contrast 
to earlier claims in the literature.
\end{abstract}
\pacs{PACS numbers: 12.38.Gc, 11.15.Ha, 11.30.Rd} }

One of the most remarkable features of the QCD phase transition is the 
simultaneous vanishing of confinement and the restoration of chiral symmetry.
Although there is much debate about the underlying mechanism which links 
the two 
transitions, their coincidence at a common critical temperature is considered 
a well established fact in lattice simulations (see e.g.~\cite{karsch} for
a review). 
Interestingly, QCD in the  
quenched approximation has an additional Z$_3$ symmetry 
which allows us to obtain additional information. For pure gauge theory
the phase transition can be described as spontaneous breaking of
the Z$_3$ symmetry \cite{Z3}. For temperatures larger than $T_c$ the
Polyakov loop acquires a non-vanishing expectation value with
a phase in one of the three sectors.
Knowing whether the observed strict correlation 
between the chiral and deconfinement phase transitions persists
in all Z$_3$ sectors could provide important clues for the understanding of 
confinement.    

In \cite{ChCh95} it was claimed on the basis of lattice calculations with 
staggered fermions, that the restoration
of chiral symmetry happens at different temperatures in the real 
($\varphi \sim 0$) and complex sector of the Polyakov loop
($\varphi \sim \pm 2\pi/3$). 
Several subsequent articles \cite{subsequent}
analyzed possible mechanisms to explain this observation.

In this letter we reexamine this problem within lattice QCD using  
chirally improved fermions. In contrast to \cite{ChCh95} we 
find that the critical $\beta$ of the 
chiral phase transition does not depend on the Z$_3$ sector,
but is coincident with the Z$_3$ breaking transition
in all three sectors.

Let us note that even if it were irrelevant for full QCD the center 
symmetry could
still lead to fascinating  phenomenological consequences in
supersymmetric Yang-Mills theories \cite{Wiese}.
Thus the investigation of its properties is also relevant 
on more general grounds.  
 
Instead of directly measuring the chiral condensate we analyze in detail 
the spectrum of the lattice Dirac operator. The density $\rho(\lambda)$
of the eigenvalues $\lambda$ of the Dirac operator is connected to the 
chiral condensate via the Banks-Casher formula \cite{BaCa80},
\begin{equation}
\langle \bar \psi \psi \rangle \; = \; - \frac{\pi}{V} \,
\rho(0) \; ,
 \label{baca} 
\end{equation}
where $\rho(0)$ is the eigenvalue density near the origin and $V$ is the 
volume of the box.
Note that exact zero-modes which come from isolated 
instantons do not contribute to the density $\rho(0)$ at the origin.
The reason is that the number of zero-modes is believed to scale as
$V^{1/2}$ and thus they do not contribute when performing
the thermodynamic limit in Eq.~(\ref{baca}). 
At low temperatures when QCD is in the chirally broken phase the density 
is non-zero at the origin, while in the high temperature phase $\rho$
 is zero in a finite region around the origin, i.e.~the spectrum
develops a gap (up to isolated zero modes) and the chiral condensate vanishes
(compare Fig.~\ref{spectra} below).
The question whether chiral symmetry is restored at the same critical 
temperature in all sectors of the Polyakov loop
can now be reformulated in terms of the spectral gap: As we 
increase the temperature, does the gap open up at the same temperature   
for all three sectors of the Polyakov loop?

Before the advent of chirally symmetric formulations for the lattice Dirac 
operator such a study was quite awkward. In particular the spectrum 
of the Wilson lattice
Dirac operator shows large fluctuations close to  the origin 
\cite{GaHi98} and
the notion of a spectral density is not well defined. The situation has
changed, since the re-discovery of the Ginsparg-Wilson equation \cite{GiWi82}.
Dirac operators $D$ which obey the Ginsparg-Wilson equation 
have eigenvalues which lie on a circle
and it is straightforward to identify a spectral density and study the 
emergence of the spectral gap. However, the only exact solution of the 
Ginsparg-Wilson equation, the overlap operator \cite{overlap}, 
has the drawback 
of being very expensive in a numerical implementation. 

Here we work with the {\sl chirally improved operator} which is a 
systematic expansion of a solution of the Ginsparg-Wilson 
equation \cite{Ga00}. In particular we use 
an approximation which has 
19 terms in the expansion and is described in detail in 
\cite{GaHiLa01}. The computation of the eigenvalues of the
Dirac operator was done with the implicitly restarted Arnoldi method
\cite{arnoldi}. 

For our quenched gauge configurations we use the L\"uscher-Weisz action 
\cite{LuWeact}. 
We work on lattices of size $L_T \times L^3$ with the temporal
extent $L_T = 6$ and two values for the spatial extent, 
$L = 16$ and $L = 20$. We use periodic boundary conditions for 
the gauge fields, while for the fermions the boundary conditions are 
periodic only for the space directions but anti-periodic for the
time direction. Our statistics is 800 configurations for the $6 \times 16^3$
lattices and 400 for the $6 \times 20^3$ lattices. We use 6 different 
values of the inverse coupling $\beta$ which gives rise to ensembles on
both sides of the phase transition. In Table~\ref{parameters} we 
list our values of $\beta$, the lattice spacing $a$ \cite{GaHoSch01}
and the temperature $T$.
We used the coefficients given by tadpole-improved perturbation theory
\cite{LuWeact,Aletal}. 
Our values for the couplings $\beta_{rt}$ and $\beta_{pg}$
for the rectangle and parallelogram terms in the L\"uscher-Weisz 
action can be found in \cite{GaHoSch01}.
\begin{table}[h]
\begin{center}
\begin{tabular}{c|cccccc}
$\beta $      & 8.10  & 8.20  & 8.25  & 8.30  & 8.45  & 8.60  \\
\hline
$a \;$  [fm]  & 0.125 & 0.115 & 0.110 & 0.106 & 0.094 & 0.084 \\ 
$T \;$  [MeV] & 264   & 287   & 299   & 311   & 350   & 391   \\
\end{tabular}
\end{center}
\caption{Parameters for our gauge field configurations.
We list the values of $\beta$, the lattice spacing $a$ 
and the temperature $T$.
\label{parameters}}
\end{table}

Let us begin the discussion of the spectral gap with a look at typical 
spectra of our Dirac operator. In Fig.~\ref{spectra} we show the
distribution in the complex plane of the 50 smallest eigenvalues $\lambda$ 
for three different gauge configurations on $6 \times 20^3$ lattices. 
The symbols are our numerical 
results and the full curve is the so-called Ginsparg-Wilson circle, i.e.~the
circle of radius 1 in the complex plane with center 1. For
our approximate Ginsparg-Wilson operator the eigenvalues do not fall exactly 
on the circle but show small fluctuations around the circle. However,  
the eigenvalues are sufficiently well ordered to allow for
the notion of a spectral density and a clear identification 
of the spectral gap. 

The plot on the left-hand side shows the spectrum for a configuration 
in the low temperature, chirally broken phase. For this case the
eigenvalues extend all the way to the origin and there is a non-vanishing 
$\rho(0)$ such that  the Banks-Casher 
relation (\ref{baca}) gives rise to a non-vanishing chiral condensate. 

The central and the right-hand side plots show spectra for configurations in 
the high-temperature, 
chirally symmetric phase. The central plot is for a 
configuration with complex Polyakov loop $P$, 
while the right-hand side result is for real Polyakov loop.
Both of these plots have a well pronounced spectral gap. The
spectral density at the origin vanishes and so does the chiral condensate.  
For the complex sector the gap is considerably smaller
than for the real sector. 
\begin{figure}[t]
\centerline{\epsfig{file=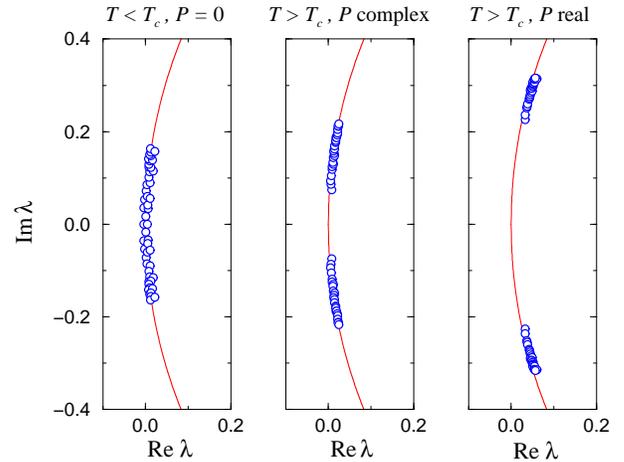,width=8.0cm}}
\caption{ Three typical spectra of the chirally improved operator.
Only the 50 eigenvalues closest to the origin are plotted: 
Left plot: the chirally broken phase 
($6 \times 20^3, \beta = 8.10$);
Central plot:
the symmetric phase ($6 \times 20^3, \beta = 8.60$) for
complex Polyakov loop;
Right plot: the symmetric phase for real 
Polyakov loop. The full curve is the Ginsparg-Wilson circle.   
\label{spectra}}
\end{figure}
 
One can understand this difference between the sectors by considering
the fermion boundary conditions. In the real sector the boundary
condition 
\begin{equation} 
\psi(\vec{x},t + 1/T) \; = \; - \, \psi(\vec{x},t) 
\label{bc1}
\end{equation} 
gives a Matsubara frequency $\pi T$ to the fermions. In the complex
Z$_3$ sectors the boundary condition is effectively 
\begin{equation} 
\psi(\vec{x},t + 1/T) \; = \; e^{\pm i \pi/3} \psi(\vec{x},t), 
\label{bc2}
\end{equation} 
giving a Matsubara frequency $\pi T/3$. In the free-field case 
(i.e.~$\beta \to \infty$) the smallest eigenvalue is equal to the 
Matsubara frequency, giving a gap 3 times larger in the case of 
(\ref{bc1}) compared with  (\ref{bc2}).   It is thus reasonable 
that the real sector gap is considerably larger than the 
complex sector gap in the interacting case too. 

In quenched QCD the finite temperature phase transition appears to 
be a weak first order phase transition~\cite{Karsch}. A
first order phase transition is governed by the mixing of two phases
and the behavior of their free energies. In our particular example we have a 
low temperature phase characterized by a vanishing spectral gap and a high
temperature phase with a finite spectral gap. For temperatures sufficiently
below or above the critical temperature the system is in only
one of the two phases while near the critical temperature the system 
shows mixing of the two phases. 

\begin{figure}[t]
\centerline{\epsfig{file=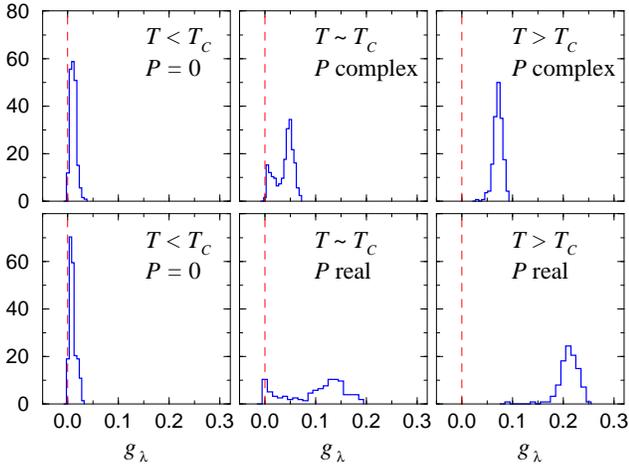,width=8.3cm}}
\caption{ Histograms for the spectral gap $a g_\lambda$. We show our results 
for lattice size $6 \times 20^3$ at three temperatures
$T < T_c$ ($\beta = 8.10$), $T \sim T_c$ ($\beta = 8.30$)
and $T > T_c$ ($\beta = 8.60$). The top row
displays the results for the complex sector of the Polyakov loop $P$, 
while the bottom row is for real Polyakov loop.
\label{histograms}}
\end{figure}

We demonstrate this mixing in Fig.~\ref{histograms} where we show histograms 
for the distribution of the spectral gap at three different values of the 
temperature. We define the spectral gap $g_\lambda$ 
to be the imaginary part of the
smallest eigenvalue which is not a zero-mode (as remarked above, zero-modes 
do not contribute to the spectral density). 
Zero-modes can be identified uniquely since for our chirally improved operator
it can be shown that they have exactly
vanishing imaginary part and the corresponding
eigenstates have a non-vanishing matrix element with $\gamma_5$, while this
matrix element vanishes identically for non-zero-modes.

We show histograms for the 
distribution of $g_\lambda$ 
for $T < T_c, T \sim T_c$ and for $T > T_c$. The top row
displays the results for the complex sector of the Polyakov loop $P$, while
the bottom row is for real Polkyakov loop. The data were computed on  
$6 \times 20^3$ lattices. Since for the real sector ($\varphi \sim 0$)
the statistics is only half of the statistics for the complex sector 
($\varphi \sim + 2\pi/3$ and $-2\pi/3$) we doubled the bin size
for the two histograms in the real sector at $T \sim T_c$ and $T > T_c$.

At $T < T_c$ we find for all sectors a single peak near the origin. This peak 
is not located exactly at 0 since also in the chirally broken phase the
Dirac operator of a finite system has a microscopical gap which 
vanishes as $L^{-3}$ \cite{randommatrix}. For temperatures near $T_c$
the histograms show a clear double peak structure characteristic for
the first order transition. The left peak corresponds to the   
chirally broken phase with a vanishing gap and the right peak is 
from the chirally symmetric phase with non-vanishing gap. As one increases
the temperature further, only the right peak 
survives. As already noted in the discussion of
Fig.~\ref{spectra} the gap is larger in the real sector, i.e. the
right peak sits at larger values of $g_\lambda$ for the real sector. 
In addition this peak is wider than the corresponding peak in the complex 
sector, i.e.~the gap fluctuates more strongly around its mean 
value in the real sector.   

In order to describe the first order transition we use a simple ansatz for 
the behavior of observables. Let us first discuss the somewhat simpler
case of the Polyakov loop. In an infinite system the Polyakov loop 
$P$ vanishes below $\beta_c$ and has a non-vanishing modulus $|P|$ 
above $\beta_c$. On a finite lattice the Polyakov loop does not vanish
exactly below $\beta_c$ but disappears like $V^{-1/2}$, i.e.~like
 $c L^{-3/2}$ with some constant $c$. Above 
$\beta_c$ the dependence on $L$ is negligible and to leading order 
$P$ is linear in $\beta$, i.e.~described by $d + k (\beta - \beta_c)$. 
Following the ideas in \cite{Borgs}
one arrives at the conclusion that near the transition the expectation value 
of the modulus of the Polyakov loop should be given by:
\begin{equation}
\langle |P| \rangle = \frac{ c L^{-3/2} \, 
e^{- \Delta f L^3  (\beta - \beta_c)} \, + \,
3 [ d + k (\beta - \beta_c) ] }{
e^{- \Delta f L^3  (\beta - \beta_c)} \; + \; 3 } \; .
\label{polfit}
\end{equation}
The term $-\Delta f V (\beta - \beta_c)$ is the difference in 
the free energies of the two phases. At $\beta = \beta_c$ the two 
free energies are equal while below $\beta_c$ the free energy  
of the chirally broken phase is smaller than the free energy of the
chirally symmetric phase and vice-versa above $\beta_c$. The factors 3 in 
the second terms in the numerator and denominator come from the 
three possible values for the phase of the Polyakov loop. 
\begin{figure}[t]
\centerline{\epsfig{file=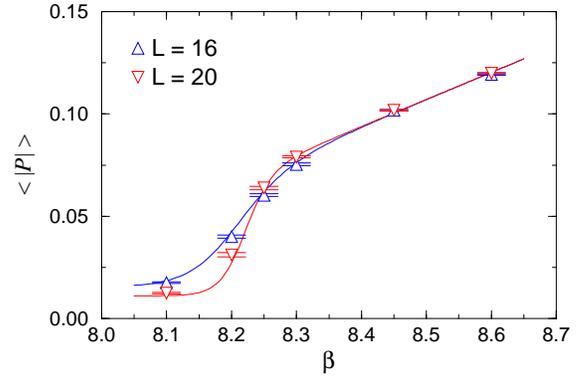,width=7.5cm}}
\caption{The expectation value $\langle |P| \rangle$ 
of the modulus of the 
Polyakov loop as a function of $\beta$. The symbols indicate 
the numerical results while the full curve is a fit to Formula (\ref{polfit}).
We display results for $6 \times 16^3$ and $6 \times 20^3$ lattices.
\label{polfig}}
\end{figure}
In Fig.~\ref{polfig} we show a fit of Formula (\ref{polfit}) 
(full curves) to our numerical data (symbols). In particular we 
present a common fit to both the $6 \times 16^3$ and $6 \times 20^3$ 
ensembles. This is possible since the parameters $d$ and $k$ are 
essentially independent of $L$. The fit result for $\beta_c$ is given
in Table~\ref{fitresults} below. The fit demonstrates that 
both the $L$ and the $\beta$ dependence are well described by
Formula (\ref{polfit}).    

\begin{figure}[t]
\centerline{\epsfig{file=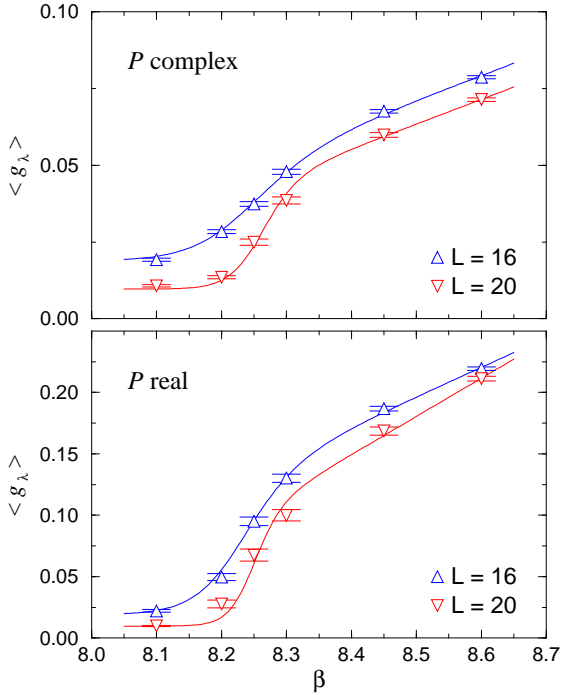,width=7.5cm}}
\vspace*{3mm}
\caption{The spectral gap in lattice units 
 as a function of $\beta$. The symbols indicate 
the numerical results and the full curve is a fit to Formula 
(\ref{gapfit}). We
display results for $6 \times 16^3$ and $6 \times 20^3$.
\label{gapfig}}
\end{figure}

For the expectation value 
of the spectral gap $g_\lambda$ we use a similar ansatz.
\begin{eqnarray}
&& \langle g_\lambda \rangle_{r,c} \; = 
\label{gapfit}
\\
&& 
\frac{ c' L^{-3} \, 
e^{- \Delta f L^3  (\beta - \beta_c)}  \, + \,  3 [ d_{r,c}(L) +
k_{r,c}(L) (\beta - \beta_c) ] }{
e^{- \Delta f L^3  (\beta - \beta_c)} \, + \, 3} .
\nonumber
\end{eqnarray}
The subscripts $r$ and $c$ indicate the real respectively complex sectors
of the Polyakov loop. 
Note that now we use the known $L^{-3}$ behavior of the microscopical 
spectral gap in the chirally broken phase
\cite{randommatrix}. In the chirally symmetric phase  
the gap is essentially linear in $\beta$ 
but the coefficients $d_{r,c}$ 
and $k_{r,c}$ turn out
to be $L$-dependent. Thus in a common fit to the $6 \times 16^3$ and
$6 \times 20^3$ ensembles these parameters had to be varied independently. 
Again the fit results for 
$\beta_c$ are given in Table~\ref{fitresults}.
In Fig.~\ref{gapfig} we plot our numerical data for the gap together 
with the curves (\ref{gapfit}). The top plot gives the results for the 
complex sector while the bottom plot shows the real sector. 
As for the Polyakov loop we find that the numerical data are 
reasonably well described by the simple first order transition formula. 

When comparing the results for the critical beta as given in Table 
\ref{fitresults} we find that within the accuracy we achieved the
spectral gap vanishes at the same $\beta_c$ for both the real and the
complex sectors of the Polyakov loop. Furthermore this value is compatible 
with $\beta_c$ as obtained from the analysis of the Polyakov loop. Combining
the three methods we find a critical temperature of $300 \pm 3$ MeV for
the L\"uscher-Weisz action which is slightly larger than the result for 
Wilson's gauge action.     

\begin{table}[ht]
\begin{center}
\begin{tabular}{c|ccc}
Measurement :   & $\langle |P| \rangle$ & 
$\langle g_\lambda \rangle_{complex}$ & 
$\langle g_\lambda \rangle_{real}$ \\  
\hline
$\beta_c$ :   & 8.24(1)       & 8.29(2)            & 8.27(2)  \\
$T_c$ [MeV] : & 296(3)        & 308(5)             & 303(5)                  
\end{tabular}
\end{center}
\caption{Values of $\beta_c$ and the critical temperature
from the analysis of the Polyakov loop and the spectral gap in 
the complex and the real sectors for the ensembles with L\"uscher-Weisz 
action and chirally improved fermions.
\label{fitresults}}
\end{table}

\newpage

\noindent
{\bf Acknowledgements:} The numerical calculations were done on the Hitachi 
SR8000 of the Leibniz Rechenzentrum in Munich. We thank the staff of the 
LRZ for training and support. C.~Gattringer acknowledges support by the 
Austrian Academy of sciences (APART 654).

\end{document}